# On a relational theory of biological systems: a natural model for complex biological behavior

PEDRO J. MIRANDA*

*Department of Physics, State University of Ponta Grossa, Paraná, Brazil*

*\*Corresponding author: pedrojemiranda@hotmail.com*

AND

GIULIANO G. LA GUARDIA

*Department of Mathematics and Statistics, State University of Ponta Grossa, Paraná, Brazil*

*Abstract:* In this paper, we develop a natural (empirical) relational theory for describing and modeling complex biological phenomena. We have as stepping stone the assertion: function implies structure. The theory is built upon a graph's theory structure in which a diffusion model of information takes place, and where dynamics can be investigated in order to generate steady quantifiers. In this context, we improve a seminal work by adding a free context biological importance measure given by the Shannon's Entropy. We also introduce the concept of biological *loci*. Such concept stands for closely related biological agents which plays a role as an agent by itself. Our results allow us to synthesize a natural model for complex biological behavior that takes into account: system's update, irreducibility, and exploit of the dynamical behavior mounted over a diffusion model. The model deals in final terms to its natural capacity to model plasticity and environmental changes, which has an intrinsic relationship with Shannon's Entropy and the sort of dynamics that biological systems can display.

*Keywords:* relational biology, graph theory, information quantification, communities, dynamics.

## 1. Introduction

Theoretical biology was formally introduced as a systematic knowledge by Nicholas Rachevsky [1]. His main contribution was to consider this subject as a dichotomist approach: metric and relational biology. The motto of metric biology consists in the following philosophy: structure implies function. From this school of reasoning, originated what it is known as Modern Biology, which encompasses most of the Biotechnology and reductionistic biology. On the other hand, the relational biology is centered on the assertion that function implies structure. For now, in order to achieve equilibrium, we assume that a general theory of biology must embrace both philosophies. Concerning the development of theoretical biology, one of the most influential researchers was Robert Rosen, whose main contributions gave the foundations to relational biology [2-4]. More specifically, he introduced the concept of systemic biology which was studied by means of a model known as $(M, R) -$ *system* [5-7]. This model was conceived by a relational view of biological phenomena exploring how its agents interact to generate its features. The present paper is developed within such view-point, *i. e.*, relational biology.

In a recent study [8], we proposed a redefinition of the importance of biological agents in a complex network, which represents a well-defined biological phenomenon. The complex network is represented by a directed graph in which agents are regarded as vertices and their interactions (*i. e.,* relations) as edges. Furthermore, we also generated a method to compute exactly such importance exploring the Perron-Frobenius properties of the transition matrix of the graph. Additionally, this procedure is applicable for any sort of biological phenomena in which its agents can be known experimentally, regardless of their nature and hierarchical level. The procedure was coined as theoretical knockout (KO), and the quantity of interest arises from the removal of a vertex and how it affects the random walk (diffusion model) within the network. This measure, the agent's importance, is computed by a relative mean error between the flux of random walkers of the original graph and the KO graph.

A wide range of investigations can be approached by this method: first because it has a strong intuitive appeal, and secondly, such method can be applied by most scientists due to its simplicity. Ultimately, it provides to experimental scientists the order of importance of biological agents, which is very handful, since most of these phenomena rely on complex relations comprised as flux diagrams (*e. g.*, metabolic pathways, trophic chains, organic systems, immunological networks, signaling networks, etc). However, we find it possible to improve our model and method in two aspects. In the first place, in some studies, it would be useful to utilize a measure that bears physical meaning with a standardized unit. Although our initial measure is useful within a fixed biological context, it cannot be used as a universal measure for other contexts since it is an index. This means that any conclusion about the agents depends only in the underlying network. Secondly, in a relational perspective, it would be also useful to calculate the importance of special sets of close related biological agents that have a functional role. For instance: closely related cells, tissues, organs, and individuals can interact in a way that it can be accounted as a biological agent by itself. Organizing the objects of improvement of our model, we list:

i. the lack of physical meaning of the previously defined measure of importance, and its context-dependent interpretation;
ii. the measure of the importance of special sets of related biological agents, and its implication in the model.

As mentioned above, these difficulties cannot be directly answered by our previous treatment [8]. Therefore, in this work, we propose an enhancement of our model including information theory to turn biological importance universal, and by introducing the concept of graph's community for defining the importance of closely related sets of biological agents.

Then we ask ourselves: how is it possible to relate biological importance to information theory? First of all, it is necessary to discuss the measure of biological importance, that is, the relative mean error, which was coined in Ref. [8]. This quantity is dimensionless and cannot be utilized to compare agents outside its context because it depends on the complex network dynamics. To surpass this problem, we propose to substitute the measure of flux to a measure of local information, namely, the well-known Shannon's Entropy [9]. To proceed in this venue, we must investigate such substitution and how it is done without loss of generality. In a broad sense, we can consider information as "the thing that gives form" from latin, *informatio*. However, in information theory, we are more concerned with the measure of information than to define it philosophically. We know that there exist many ways to measure information, but in this particular paper we shall consider the following situation: let *X* be a discrete random variable; the measure of the information of *X* is the measure of how much "effort" it takes to describe it. From these basic notions, we can derive Shannon's Entropy for a given distribution of a discrete random variable [9]. The association of the flux and the quantity of information can give us a steady measure of information based on a random walk in a directed graph.

The second difficulty previously mentioned is the importance of special sets of related biological agents. To solve this problem we apply the concept of community (from graph theory) [10]. It can be roughly defined by a set of vertices with high edge density when compared to the remaining vertices of the graph. Thus, since the community arises naturally from the structure of the network, its KO can provide a corresponding behavior in the information processing in the resulting network.

As far as such results are implemented, the model changes along with the way that we look at it. This means that some biological considerations must take place in order to turn our model coherent, intelligible, and empirically accessible. Thus, we propose a synthesis of the results to respond to such demands. In this work, we aim to bridge the biological importance to the quantity of information (*i. e.*, entropy) by means of a diffusion model [8]. We also intend to generalize such implementation to the knockout of whole communities in order to understand how it affects the information flux. Additionally, we propose the investigation of the eventual periodicity that an asymptotic state vector (*i. e.*, flux vector) may display for particular cases. In order to organize such objectives, we list:

I. To generalize the measure of biological importance in terms of information theory.
II. To implement the knockout of communities and to investigate how it affects the flux of information.
III. To synthesize the obtained results into a natural model that helps us to understand complex behavior of biological phenomena.

## 2. Mathematical development

*2.1 Review of the initial model*

The model starts with a directed graph constructed by means of an experimental or observational biological data set, which associates biological agents (*i. e.*, molecules, cells, tissues, individuals, etc) to vertices; and its interactions to directed edges [8]. Given such a graph, we can extract its associated adjacency matrix, which in turn can give us the transition matrix based on the out degree of all vertices. Once the graph is built, we submit it to a process of a random walk, in which all zero in-degree vertices receive a direct edge from an origin. The origin is a vertex that generates walkers in the graph in each time step *t*. At each time *t*, a walker is created in the origin and shift position once from vertex to vertex, and all previously created walker also shift positions. To shift positions, it has to pass through edges respecting its directionality. As times goes on, the number of walkers raise over time, and the system (*i. e.*, graph or network) will have a distribution of walkers throughout its vertices. If we denote by $\sigma_i(t)$ the number of walkers in a vertex *i* in the time *t*, then the flux on this vertex is the relative number of walkers on it; formally $f_i(t) = \frac{\sigma_i(t)}{N(t)}$, where $N(t) = t$ is the total number of walkers in the time $t$ [8]. This summarizes our model of a directed graph in which a stochastic dynamics plays a role by means of a random walk that respects the direction of edges.

With this notation, we can define the flux vector of the system in the time $t$ and $n$ vertices, which actually also stands for the state vector of our system's dynamics,

$$\boldsymbol{F}(t) = \big(f_1(t), f_2(t), f_3(t), \dots, f_n(t)\big). \tag{1}$$

We can now consider that our initial directed graph has suffered a vertex removal, then its adjacency matrix changes, and this lead to a change in the transition matrix. This implies that the dynamics of the random walk upon the graph has been altered and the flux vector changed from $\boldsymbol{F}(t)$ to $\boldsymbol{F}'(t)$. With these state vectors, one for the whole graph and other for the modified graph, we can compute the distance between its components, *id est*

$$\boldsymbol{D}_{G,G'} = \boldsymbol{F}(t \to \infty) - \boldsymbol{F}'(t \to \infty) = (\Delta f_1, \Delta f_2, \Delta f_3, \dots, \Delta f_n = f_n), \tag{2}$$

where $\Delta f_i = \lim_{t \to \infty}[f_i(t) - f'_i(t)]$, for $1 \leq i \leq n$. Since the values of $\Delta f_i$ can be positive or negative, we define a third local measure $\mu_i$ which scales which the tendency of increasing flux or decreasing flux, respectively. Formally, we have

$$\mu_i = \begin{cases} \dfrac{\Delta f_i}{f_i(t \to \infty)}, & for\ \Delta f_i > 0 \\ \dfrac{|\Delta f_i|}{f'_i(t \to \infty)}, & for\ \Delta f_i < 0. \end{cases} \tag{3}$$

The value of $\mu_i$ is called *relative error*, and stands for the local impact upon the removal of the vertex in the modified graph. Having the collection of all relative errors, the quantifier for biological importance is defined by $\mathrm{M}_{G,G'} = \frac{\sum_{i=1}^{n} \mu_i}{n}$ for all $1 \leq i \leq n$, , where $G$ is the original graph that generates the state vector $\boldsymbol{F}(t)$ and $G'$ is the modified (knocked out) graph that generates the state vector $\boldsymbol{F}'(t)$.

In our previous work [8], we have proposed two methods to compute $\mathrm{M}_{G,G'}$: one based on an algorithm and another based on algebraic results. Here, we will only consider results of the second type, since the computation does not fit into our proposal.

In order to calculate $\mathrm{M}_{G,G'}$, we need to evaluate the vectors $\boldsymbol{F}(t)$ and $\boldsymbol{F}'(t)$ when $t \to \infty$. As we emphasized above, any graph has an associated adjacency matrix, thus transition matrix, also known as probability matrix [11]. Since the probability of walker to access a given vertex $i$ depends only on its out degree, the entries of the probability matrix $T$ is given by $p_{ij} = \frac{1}{k_{i\,out}}$, where $k_{i\,out}$ is the out degree of vertex $i$.

If we have a state vector at any initial condition $F(t = 0)$, the following operation is applicable

$$\boldsymbol{F}(t = 1) = T\boldsymbol{F}(t = 0) \Rightarrow$$

$$\boldsymbol{F}(t = t_\ell) = T^{t_\ell}\boldsymbol{F}(t = 0). \tag{4}$$

As it has been stated before, we are interested in the flux vector when $t \to \infty$, which implies that it is necessary to calculate $T^\infty$ in order to find such a vector. The numerical calculation of $T^\infty$ is inconvenient, because it demands too much computational time and effort. However, there is an alternative and convenient way to compute the flux vector when $t \to \infty$, and it is based on the Perron-Frobenius features of the transition matrix of the subjacent graph in which the random walk is performed [11]. Let $\{\lambda_i\}$ be the set of eigenvalues of the transition matrix $T$ of a given directed graph. The Perron-Frobenius features will hold if, and only if, $|\lambda_1| \geq |\lambda_2| \geq |\lambda_3| \dots \geq |\lambda_n|$ and $|\lambda_1| = 1$. If these criteria are satisfied, then the flux vector $\boldsymbol{F}(t \to \infty)$ exists as a unique asymptotic state [10]. The computation of such a vector is simple and has been performed at Ref. [8].

*2.2 Information theory: translation from flux vector to Shannon's Entropy*

As mentioned above, in the model given previously, we introduced the relative mean error as a measure of biological importance. Since it is an index, then it is dimensionless; this implies that it bears little physical meaning. To improve this initial model, in this paper, we change that measure to the well-known Shannon's Entropy [9], which has physical meaning. To proceed with this translation, let us recall the set of the flux vector's components $\{f_i(t \to \infty)\}$, for all $1 \leq i \leq n$. This set can be understood as a probability distribution. If the Perron-Frobenius criteria hold, then there exists the asymptotic state vector $\boldsymbol{F}(t \to \infty)$, which implies that such distribution is also unique. Thus, it is natural to define the quantity of information by means of Shannon's Entropy [9] applied to the flux distribution $\{f_i(t \to \infty)\}$ as

$$H = -\sum_{i=1}^{n} f_i(t \to \infty) \ln f_i(t \to \infty). \tag{5}$$

Shannon's Entropy is an extensive property and measures the information content that is associated with the flux vector in the asymptotic state. Its unit is given by *nats* by *channels*. This application is convenient since it can provide links to the quantification of other thermodynamic variables such as internal energy and temperature. However, it is important to understand why this translation is possible and convenient. In his work, Shannon has defined entropy in terms of a discrete random variable *X* that can have values within a given set $\{x_1, \dots, x_n\}$ which comprises a probability mass function $P(X)$. If we denote *E* the expected value operator and *I* the information content of *X*, we have

$$H(X) = E[I(X)] \Rightarrow$$

$$H(X) = E\big[-\ln\big(P(X)\big)\big], \tag{6}$$

where $I(X)$ is, by itself, a random variable. The Eq. (6) can be rewritten in the following way

$$H(X) = -\sum_{i=1}^{n} p(x_i) \ln p(x_i). \tag{7}$$

In the cases where $p(x_i) = 0$, we define

$$\lim_{p \to 0+} p_i \log p_i = 0. \tag{8}$$

With this reasoning in mind, we are able to define the new measure of biological importance. Let us recall that the importance of agent was due to a knockout of one vertex that generates a KO graph *G'*. It is clear that such a graph allows us to find $\boldsymbol{F'}(t \to \infty)$ since that Perron-Frobenius criteria always hold in our model. Thus, computing the Shannon's Entropy of this vector, we define the variation of the corresponding entropies as the new measure of biological importance, *id est*

$$\Delta H = H_G - H_{G'}. \tag{9}$$

If $\Delta H$ has negative value, the topological change made in $G'$ implied in an increase of entropy, otherwise implies a decrease. The biological interpretation of such result will be approached in Subsection 3.1.

*2.3 Community analysis: from vertex removal to community removal*

We have already worked on the knockout (KO) of single vertices and also provided a way to interpret its relevance according to its biological context [8]. Now, we argue about the relevance of special sets of close related vertices. In order to define such sets, we utilize the definition of communities applied in terms of graph theory. Rough speaking, a community can be regarded as a set of vertices that have a high edge density between them and low edge density to the rest of the graph. Although it is not easy to define community formally, since the available definitions are restrictive or computationally inefficient, we utilize the definition shown in [11] due to its consensus with the literature. Such definition is given in the sequel:

**Definition 1.** *Let $G = (V, E)$ a finite graph. Consider that $\mathcal{P} = \{C_1, \dots, C_l\}$ is a partition of the set of vertices V of G (i. e., $\bigcup_{i=1}^{l} C_i = V$, where $C_i \cap C_j = \emptyset$, $\forall i \neq j \in \{1, 2, \dots, l\}$ ). One says that $\mathcal{P}$ is an efficient representation for the community structure if the proportion of edges inside each $C_i$ (internal edges) is high when compared to the proportion of edges between them. These subsets $C_i$ are called communities.*

At this point, it is important to emphasize the difference between the concept of community in the general sense, and the concept of "biological community". Seeking for convenience, and for a clear understanding of this work, we introduce the term of *biological locus*, or the contraction *bio-locus*. This term refers to a community that comes from a biological phenomenon. We propose this differentiation because the term community arose from social complex network studies, which has a totally different meaning. In analogy, we denote the term for community structure by *bio-loci*.

The next step is to know how to find the bio-loci, and consequently, all bio-locuses of a graph derived from a biological phenomenon. There are some methods to compute such sets. We strongly recommend, due to precision and applicability, the *Walktrap Algorithm* (WTA) [10]. This method is widely utilized in the computational applications of graph theory. However, since we are not interested in the method *per se*, we will not extend ourselves on such endeavor.

Thus, for simplicity, let us consider that we have found the bio-loci of a directed graph that represents a biological phenomenon. In other words, we know all bio-locuses and their corresponding vertices. So we can generalize the biological importance to the bio-locus. This is performed by knocking it out from the graph, that is, removing all its vertices at the same time. The procedure of calculation is totally similar to the vertex removal. Thus, one has a value of importance to the biolocus $M_{G,G'}$, which can be translated to the Shannon's Entropy variation, as defined in the previous section.

## 3. Biological implication of the model

*3.1 Flux's Entropy: the way that biological systems update itself*

Provided some environmental constraints, it is well-known that biological systems tend to steady states in which the entropy is minimal, or, equivalently, to states of minimal free energy [12]. In Subsection 2.2, we have defined Shannon's Entropy for a given flux vector (*i. e.*, a state vector), that now we will call Flux's Entropy. This means that for each state vector there will be a value for entropy. It is clear that the set of state vectors and its associated values of entropy behave as a degenerated system. This means that the system can reach a value of entropy by many different ways by updating its dynamics [13]. We are taking as premise *autopoiesis* as a core phenomenon that drives biological complex behavior [14]. An autopoietic system is one of a kind that builds itself to a limit; such limit is given by environmental forces that act on it and by its ability to change internally. This is equivalent to say that the system contains the possible responses to its necessity according to the environment, but not determined by it.

Let us synthesize what we have defended until now: biological phenomena are open systems that tend to steady states of minimal free energy provided some environmental constraints. As far as such constraints change, the system tends to new steady states; such plasticity to adapt internally to new constraints is described by autopoiesis. This implies that the system takes control over its behavior since it is a degenerated system, the same thermodynamic state can be achieved by different system's configurations.

A natural question is posed in this context: how can a system generate itself? Since we are dealing with autopoietic systems, the system is driven ultimately by itself. Thus, to address such questioning, we make use of the concept of emergent property: in a system composed of perceivable parts, an emergent property is one of a kind that arises in a non-deductible way from its parts. This means that the driving force that exactly determines the future states of a biological system is not contained in the environment, neither in its parts, but rather in its irreducible efficient cause [15]. In any case, we will be able to calculate such irreducibility of a system in the next section.

For now, let us work on the formality of such complex behavior. In order to update its dynamics, a system must act over its internal relations, which ultimately will change the transition matrix of the given system. More precisely: when we say that the system will update its dynamics, in mathematical terms, we refer to a change in the dynamical rule that governs a biological system represented by a graph. This means that the entries of the transition matrix change accordingly to the system's demands. Such change, in terms of graph theory, occurs by putting weights on the edges, to silencing (KO) some of its vertices, or by adding a new edge or vertex. In any case, the entries of the matrix change, so its dynamics. As an example, consider the updating process displayed in Fig. 1.

It is important to note that a set of conditions, both external and internal, puts a limit on an array of states that corresponds to a minimal free energy state. One way to compute such state is to calculate one thermodynamic variable. Our model provides Flux's Entropy to this end.

We would like to conclude this section defending that the computation of Flux's Entropy is a steady and trustworthy measure of biological importance. Such quantifier has deeper physical and biological meaning, as far as it is linked to thermodynamics and its state functions. This result gives us context to start the understanding of how the system updates itself among its possible configurations. Furthermore, this quantification comes uniquely from the way that things relate to each other in a biological phenomenon. In other words, we are within our major relational premise: function implies structure.

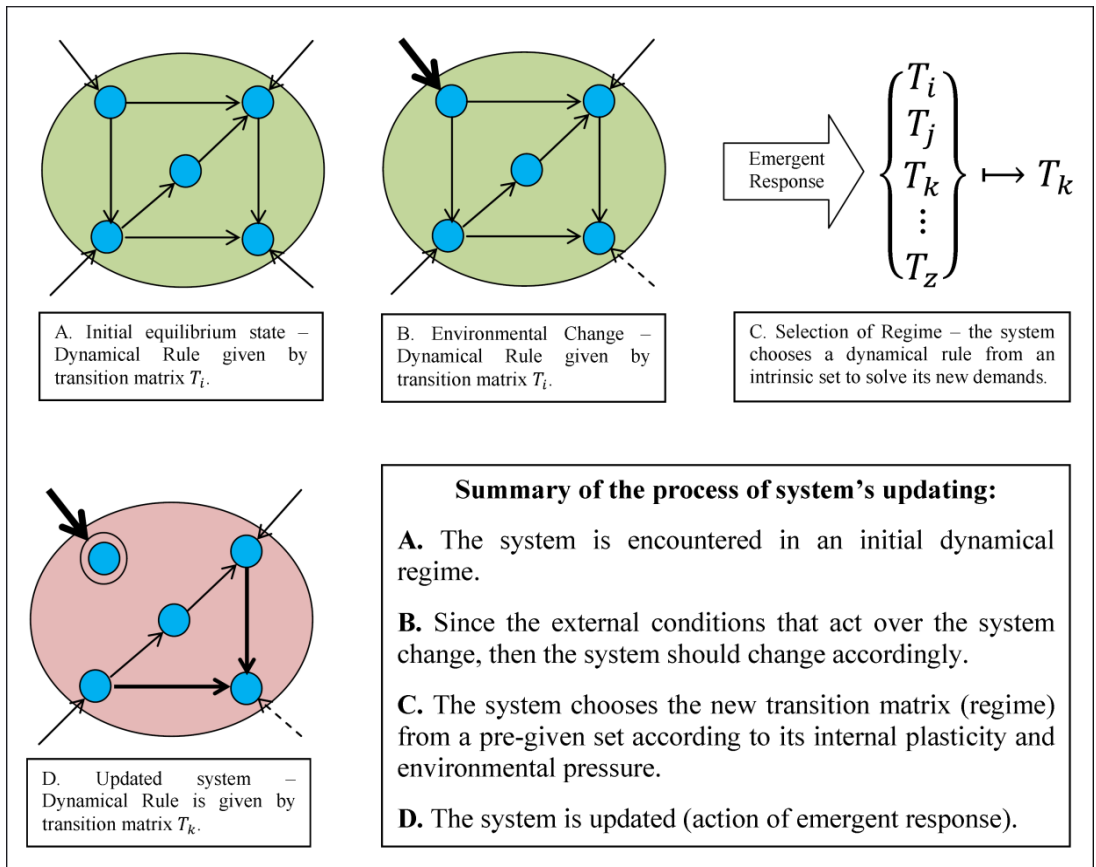

Fig. 1. An example of system's update processing.

*3.2 Biological loci: revisiting the concept of Rosen's coarse structure*

    To interpret the concept of bioloci introduced in this model, let us take some examples of how such concept arises naturally from living phenomena. In the cellular level, for instance, we can consider sets of cells that have a well-defined role, such as tissue glands, neuron complexes, muscles fibers, pulmonary alveoli, neural ganglia, etc. In other words, every cell association can be considered as a functional unit in which, in our model, stands for the community structure. Considering hierarchical level, we can regard sets of tissue that compose organs, such as the skin layers, the layers of muscle that envelop the stomach, the bone composition of the hard and soft tissue, and so on. If we take individuals as hierarchical level, we can consider biological populations of individuals that share the same space and live at the same time, or we can also regard individuals that are included in the same generation, or even the set of individuals regardless of their species that share a common niche.

    In every given example, we can regard them as a unity composed of parts that behave according to its unity. Thus, the criteria for defining such special sets, in a biological perspective, are the unity based on functions. At this point, we will revisit a concept introduced by Rosen in his seminal work on $(M, R) - system$ [5]. The concept in question is the coarse structure that refers to abstract systems taken as "black boxes". By definition, in the coarse structure, one only knows its input and output materials, but not its internal operations. Rosen utilized such concept due to his interest to organize a representational system for biological phenomena. However, we are interested in the knowledge of the internal operations of the coarse structure. Because of this, we propose a conceptual substitution of coarse structure to the *bioloci* defined in a former section.

    The theoretical gain in such substitution is the fact that it arises naturally in a system composed of parts, and we also know how it works. Since bioloci is a graph, then we know its vertices and edges, so we can understand its internal operations. Additionally, we know its local dynamical rule from its transition matrix. As the bioloci presents the internal structure of biologic agents, let us quantify the loss of information without considering such internal structure.

    To address such problem, we will introduce a measure of complexity for an observable *biolocus* found by means of Def. 1 and the WTA [10]. Thus, let us suppose that we have computed such structure from an initial graph, and we want to know if its information processing is relevant enough that it cannot be replaced by a single vertex. To support our approach we explore the process described in Fig. 2.

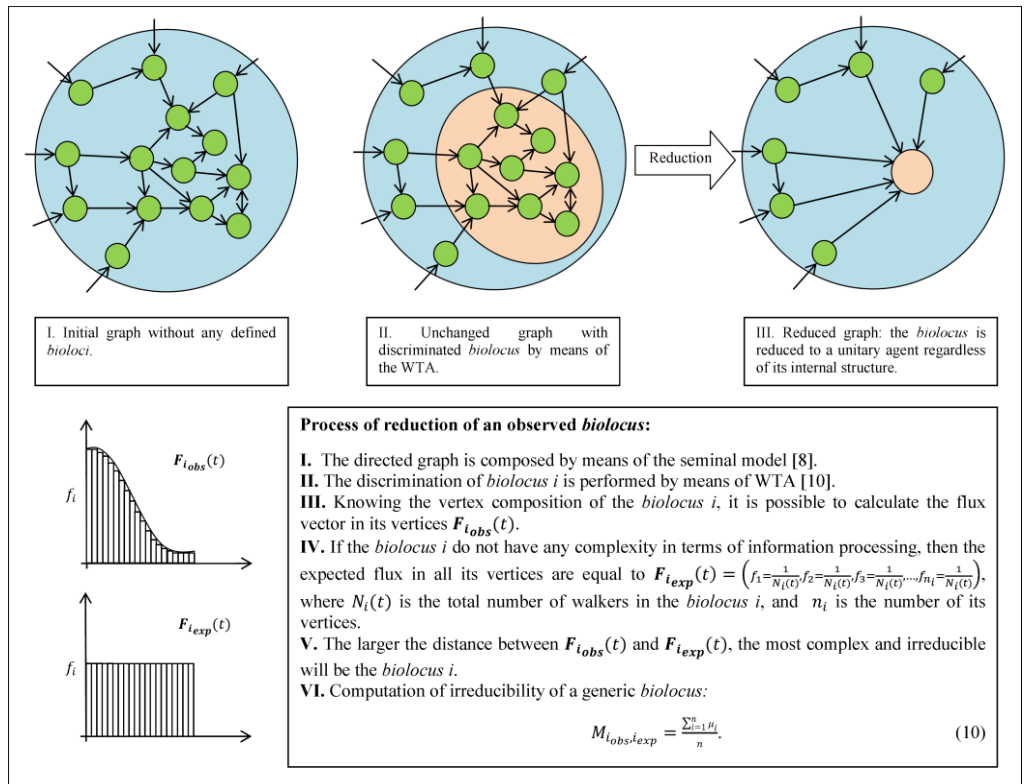

Fig. 2. Schematics of the reduction of *biolocus*.

The directed graph presented in Fig. 2 (I) is what Rosen addressed as *fine structure* [3], in which all operations are known in details. The discrimination process Fig. 2 (II) allows us to known parts of the system that behave as wholes. If such whole could be reduced by its parts, then we expect to have a uniform distribution as expressed in the state vector $F_{i_{exp}}(t)$. A real and observed biolocus has a state vector $F_{i_{obs}}(t)$. The relative mean error (Eq. 10) is then adapted to compute the overall separation between such vectors. The values of $M_{i_{obs},i_{exp}}$ closer to zero implies that the observed biolocus bear little complex behavior in terms of information processing, and then its reduction to a single vertex generates little impact. Otherwise, if $M_{i_{obs},i_{exp}}$ is close to one, then the biolocus has relevant information processing that would be lost if reduced to a single vertex. All discussions are based on the role that the parts take as far as it behaves as a whole. Next, we will discuss how the system may utilize such feature in its own *poiesis*.

3.3 Dynamics on biological system: plasticity plus environmental constraints

Change is an omnipresent aspect of biological phenomena, and this aspect is an almost synonym of dynamics. Actually, biological phenomena are maintained by their capability to respond promptly to external *stimuli*. Such capability is denominated plasticity. Rough speaking, biological dynamics is cornered between its plasticity and its environmental constraints. This simplification implies that if the system plasticity and the environmental constraints imposed on it are known, then its behavior is also known. As matter of fact, this is not what happens, and we shall see the reason by means of our model.

Plasticity is the capability of a system to change due to a stimulus (*i. e.*, environmental change). In other words, the capability of changing the directed graph's structure, ergo its transition matrix. However, there exists an association between plasticity and environmental change: each environmental change corresponds to a set of possible system's configuration, that is, a set of transition matrices (see Fig. 1). The bigger the set, the bigger will be its plasticity to a particular stimulus. On the other hand, environmental changes stand for the set of inputs and output of a particular system; in other words, the relationships that the system has with its vicinity. But how does a system choose between these arrays of matrices?

The answer to this question lies upon the regime that the system is submitted. As a major example, consider systems whose work on the homeostatic regime. Homeostasis is a property of a system in which its internal variables are kept within a very fine range, almost constant. In terms of dynamics, this is the case of an asymptotic stationary behavior. We are considering the asymptotic behavior because it is built upon internal self-regulations that come with sufficient time. Note that any external change or internal malfunctioning can make the system lose its stationary behavior. In such case (homeostasis), the system modulates itself in order to return to its characteristic stationary behavior. Although we do not address these problems (non-stationary behaviors) here, we will work in this subject in a forthcoming paper.

At this point, we will consider the concept of entropy as the main role in our model. When the system is dragged from an initial state, it must respond to this deviation by choosing a transition matrix to fix itself. The process of selection must involve the premise that biological systems tend to steady states in which the entropy is minimal [12]. However, for each value of entropy, there will be a subset of state vectors in which the system updates itself. As far our model provides the computation of entropy, we know the subset of vectors with the corresponding value of entropy. Since Shannon's Entropy rely only upon the flux distribution, the number of state vectors with same entropy is given by the multiplicity of states [16],

$$\Phi(t) = \frac{N(t)!}{\sigma_1(t)!\, \sigma_2(t)! \ldots \sigma_n(t)!}. \tag{14}$$

Ultimately, the system's criterion for choosing a dynamical state is based on its corresponding entropy.

Another major example of behavior is homeorhesis, *i. e.*, systems having natural periodicity. Homeorhesis is the property of a system in which its dynamical orbits are bounded within some attracting basin. This property is mainly observed in ecological systems. When such system is dragged from its natural periodicity, then it must update itself in the same manner of the former case. However, the criterion now is to be periodic; more precisely, to bear a particular periodicity. In this case, the entropy is associated to each periodic state vector. For the natural functioning of the system, it is necessary to have periodic checkpoints, in which each checkpoint is associated with a particular value of entropy that is natural to the system. In other words, the system must reach a certain value of entropy periodically. Other examples of natural periodic behavior stand for the heart beats (in normal conditions), the ventilation rhythm of lungs, the rate of triggering of excitable membranes, and so on. As stated above, this approach will also be included in a future paper.

Here, we give one example of a system that is naturally chaotic: neural networks of the brain. In this case, when the system is dragged from its natural chaoticness, in a similar way to the former cases, it must compensate itself. It is important to notice that in any case, the system has a limit to its capability to update itself, and such capability changes in the lifetime of the system. For instance, in an individual, its age corresponds to its capability to deal with internal and external modification; as times goes on, this capability decreases. On the other hand, in ecosystem level, the older the system, the abler it is to respond to external perturbations since its internal interactions are stable and fixed (*i. e.*, climax). It induces us to conclude that the response capability of a system depends mainly on its nature.

## 4. Final Remarks

In this paper, we have introduced a new measure of biological importance in terms of Shannon's Entropy, namely, the Flux's Entropy. We also generalized the concept of communities applied to biological systems. Such generalization allows us to compute its importance as a biological agent *per se*. The new notion of biological community, *i. e.,* biolocus, brought the subject of irreducibility. This concept is quantified by means of the relative means error in terms of a uniform distribution. All these results induced us to synthesize our main results into a relational theory of biological phenomena naturally. As future works, we intend to investigate specifically nonlinear dynamics: periodic and chaotic behaviors.